# Logistic Map Pseudo Random Number Generator in FPGA

Mateo Jalen Andrew Calderon, Lee Jun Lei Lucas, Syarifuddin Azhar Bin Rosli, Stephanie See Hui Ying,
Jarell Lim En Yu, Maoyang Xiang, T. Hui Teo*
Engineering Product Development
Singapore University of Technology and Design
*Corresponding Author: tthui@sutd.edu.sg
These authors contributed equally to this work.

**ABSTRACT**

This project develops a pseudo-random number generator (PRNG) using the logistic map, implemented in Verilog HDL on an FPGA and processes its output through a Central Limit Theorem (CLT) function to achieve a Gaussian distribution. The system integrates additional FPGA modules for real-time interaction and visualisation, including a clock generator, UART interface, XADC, and a 7-segment display driver. These components facilitate the direct display of PRNG values on the FPGA and the transmission of data to a laptop for histogram analysis, verifying the Gaussian nature of the output. This approach demonstrates the practical application of chaotic systems for generating Gaussian-distributed pseudo-random numbers in digital hardware, highlighting the logistic map's potential in PRNG design.

## I. Introduction

Our project set out to create a reliable pseudo-random number generator (PRNG), an essential component in many technological applications today, and implement it using the CMOD-A7. Our interest was piqued by chaotic systems, known for their sensitivity to initial conditions, or "seeds," making them an intriguing choice for PRNGs. Among various chaotic models we explored, including the Lorentz equation and the double pendulum, the logistic map captured our attention due to its straightforward yet chaotic nature, positioning it as the centrepiece of our investigation.

The logistic map's appeal lies in its deterministic behaviour that, when initiated with a slight variation in initial conditions, can lead to vastly different outcomes. This sensitivity to initial conditions is a hallmark of chaotic systems and is particularly valuable in the context of PRNGs, where unpredictability is a desired feature. Inspired by this, our project aims to leverage the logistic map to develop a PRNG that is not only efficient and dependable but also implementable on a Field-Programmable Gate Array (FPGA), a platform celebrated for its versatility and capability for rapid computations. We code out the different modules we require for this implementation using Verilog HDL on Vivado, and we use a CMOD-A7 FPGA board along with a 7-segment display for testing and visualisation.

This report outlines our journey from exploring the potential of chaotic systems for PRNGs to the detailed implementation of the logistic map on an FPGA, culminating in a system that generates Gaussian-distributed pseudo-random numbers, underscored by the innovative use of a sensor-derived seed. Through this endeavour, we aim to demonstrate the practical utility and effectiveness of chaotic systems, especially the logistic map, in the realm of pseudo-random number generation.

# II. Literature Review

In exploring the development of pseudo-random number generators (PRNGs) using chaotic systems, we did a literature review to find insights from several studies on the logistic map, coupled maps, and statistical methods for achieving Gaussian distributions from chaotic outputs, [1]. Our focus was on understanding the dynamics of the logistic map, adapting the Central Limit Theorem (CLT), [2] for dependent variables, and employing the Exponentially Weighted Moving Average (EWMA), [3], technique to allow us to correctly make a pseudo-random number generator that followed a Gaussian distribution.

The logistic map function is given by,
$$x_{n+1} = r \cdot x_n (1 - x_n) \tag{1}$$

The logistic map is recognised for its simple yet chaotic behaviour, offering a rich source of unpredictability essential for PRNGs. Its sensitivity to initial conditions, a characteristic feature of chaotic systems, is particularly valuable for generating pseudo-random sequences. It works when $r$ is in the chaotic range, the value of $x$, the iterating value of $x_{n+1}$ essentially becomes non-periodic, and does not have a pattern, even though it is still deterministic. This is why we chose the logistic map function for our PRNG.

Coupled maps extend this complexity, suggesting that interactions between multiple chaotic systems can enhance the unpredictability and quality of the generated sequences. This approach aims to leverage the compounded chaos for more robust PRNGs. But in our implementation, coupled maps, or specifically coupled logistic maps would make it way too complex.

Transforming the output from chaotic systems to follow a Gaussian distribution is crucial for many applications. However, the dependency of successive outputs in chaotic systems poses a challenge to traditional methods like the CLT. Research in this area has focused on adapting the CLT to accommodate these dependencies, exploring theoretical frameworks and practical techniques to achieve normally distributed outputs from chaotic sequences.

The EWMA method is highlighted as a tool for mitigating the correlation between successive values in a sequence, facilitating the application of the CLT to sequences generated by chaotic systems. This approach helps in smoothing the data and making it more suitable for generating Gaussian-distributed pseudo-random numbers.

# III. Pseudo Random Number Generator Algorithm

After understanding more about the chaotic systems, CLT and EWMA, we tried to make tests and prototypes before implementing things in Verilog HDL.

The reason for using a chaotic system like the logistic map is mainly due to its deterministic nature and chaotic, unpredictable nature, which is sensitive to an input seed value. These make it ideal for use in a PRNG.

With respect to Equation (1), we take $R = 4$, which makes this chaotic. This works for any other $3.5 < r < 4.0$.

We use the output from the logistic map function and take its value EWMA, as this would allow for the application of CLT, such that the EWMA output is normal. We use EWMA as it is a simple, efficient method that allows the data to be averaged or aggregated so that they may exhibit properties of independence, more so than the raw outputs of the logistic map function. Based on the literature review, we saw that CLT techniques can be applied to datasets that are not exactly independent or that can be applied to moving averages. From this knowledge, we try to use EWMA so that CLT may be applicable in our case.
The EWMA function is given by,

$$EWMA_t = \alpha \cdot r_t + (1 - \alpha) \cdot EWMA_{t-1} \qquad (2)$$

In our case, we take $\alpha = 40/50$ from experimenting with the algorithm in Julia.
In fact, in many real-world applications, especially in finance and economics, EWMA and similar smoothing techniques are used to analyse time series data. For practical purposes, such as risk assessment and forecasting, the resulting smoothed series are often treated as if they were normally distributed. This is why we thought of using the EWMA to make our output conform to a normal or Gaussian distribution.

We did some proof of concepts using code, done in Julia. We found that we can indeed empirically obtain a distribution of PRNG outputs close to a Gaussian Distribution. However, whether it is indeed a Gaussian distribution or not and whether CLT can apply is something that we shall not theoretically prove in this report, as it may take attention away from our original intention. Nonetheless, we provide sources in the references that support the use of chaos theory and EWMA to make a normally distributed PRNG.

## Proof Of Concept

For this, we used the Plots package in Julia, and used a simple logistic map function to generate a random number. From here, we used the idea of EWMA to ensure that our final output value for this PRNG would be a value that would conform to a Gaussian Distribution by CLT, [4]. We double-check that it follows a Gaussian distribution by plotting a histogram of the PRNG outputs and a Gaussian Curve to overlay it. We also confirm whether the PRNG is "random" by plotting the outputs against time (iterations) in this case, as shown in Figure 1. The values observed over the iterations are not periodic and are hard to predict and determine, making it seem random to those unaware of the algorithm being used.

```julia
using Plots
maxiter = 1000 # More iterations for more data
window_size = 50 # Window size for the rolling average
function f(x, r=4); (r * x * (65535 - x) / 65535); end

x0 = 6000; stored_x = zeros(maxiter); stored_x[1] = x0

for n = 2:maxiter; stored_x[n] = f(stored_x[n-1]); end

# Calculate the rolling averages
rolling_averages = zeros(maxiter); rolling_averages[1] = stored_x[1]
for i = 2:maxiter; rolling_averages[i] = (40*rolling_averages[i-1] + 10*stored_x[i])/50; end

for i = 1:maxiter; rolling_averages[i] = floor( rolling_averages[i]); end

#Plotting the distribution of rolling averages
histogram(rolling_averages, bins=10, label="PRNG Output", xlabel="Value", ylabel="Frequency", title="Distribution of PRNG", xlims = (0,65535))
using Statistics # For mean()

# Assuming mean_val and std_dev are already calculated
mean_val = mean(rolling_averages); std_dev = std(rolling_averages)

# Manually calculate the PDF values
function normal_pdf(x, μ, σ); return (1.0 / (σ * sqrt(2 * π))) * exp(-0.5 * ((x - μ) / σ)^2); end

 # Generate the x values (ensure this covers the range of your data)
x_vals = collect(1:2:65535)

# Calculate the PDF for each x value
normal_dist = [normal_pdf(x, mean_val, std_dev) for x in x_vals]

# Overlay the manually calculated normal distribution
plot!(x_vals, normal_dist, label="Normal Distribution", linewidth=2, colour=:red)
```

```
34.
35. # Run separately this next line to show values against iter to confirm whether random or not
36. plot(1:maxiter, rolling_averages, label="PRNG Output", ylabel="value", xlabel="iter", title="Evolution of
PRNG output over time")
```

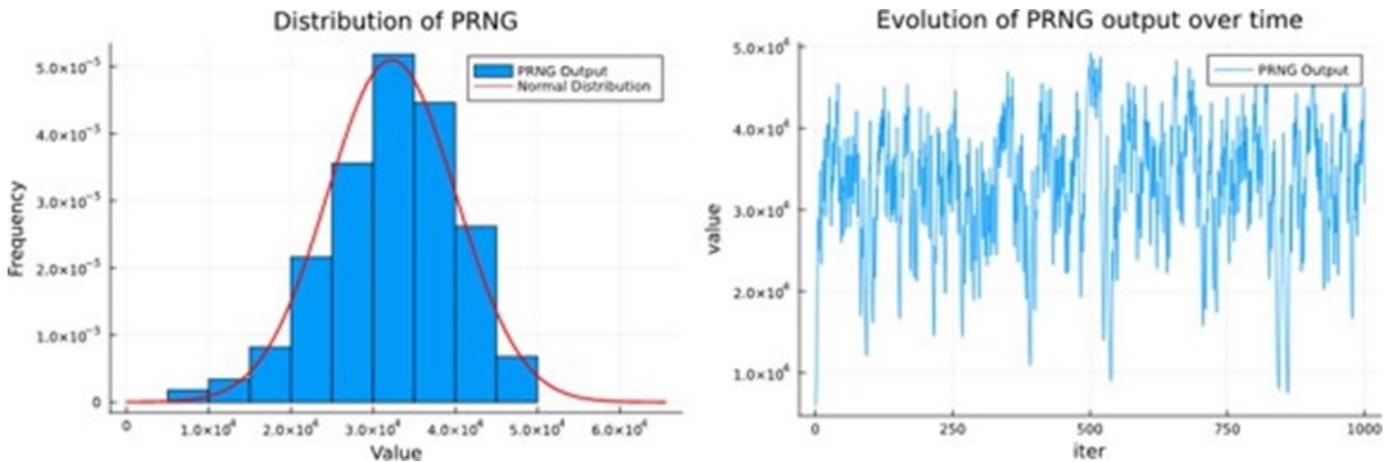

*Figure 1: Proof of Concept plots for PRNG algorithm*

# IV.   FPGA Implementation of PRNG

The big problem with using this is that, typically, in FPGA, the calculations deal with discrete numbers. Dealing with decimals would suggest the need for many shift registers, making the implementation overly complicated. As such, we rescale the original logistic map function. We cap the possible number of values we want to display at 216, the total number of values that can be represented by 16 bits.

Thus, we rescaled the logistic map function by 65,535 (as seen in the codes below) and floored each value obtained. Based on the logistic map function, this would always give us a pseudo-random integer between 0 and 65,535.

Implementing this in Verilog HDL, we created two modules: ***chaotic_Lmap*** and ***EWMA_avg***. A seed initiates the first output value, which iteratively passes through the ***chaotic_Lmap*** module. This function gives an input to EWMA, which then returns our desired pseudo-random numbers.

```verilog
1.  module chaotic_Lmap(
2.  input clock,
3.  //input reset,
4.  input [15:0]xt,
5.  input [7:0] r,
6.  output [15:0]xtnext
7.      );
8.
9.  reg [15:0] xtnext_r;
10. assign xtnext = xtnext_r;
11.
12. wire [31:0] intermediate;
13. assign intermediate = r*xt*(65535-xt);
14. //reg next;
15. //assign ctnext = next;
16. always @(posedge clock ) begin
17. //     intermediate < = r*xt*(65535-xt);
18.     xtnext_r <= (intermediate + 32767) / 65535;
19. end
20.
21. endmodule
```

```verilog
module EWMA_avg(
//input wire [15:0] current_avg,
input clock,
input [15:0] seed,
input reset,
input [15:0] xt,
output[15:0] avg
    );
wire [31:0] inter;
assign inter = (avg*40 + 10*xt)/50;
reg [31:0] seedinter;
reg [15:0] rr;

reg [15:0] avg_r;
assign avg = avg_r;

always @(posedge xt or negedge reset) begin
    if (!reset) begin
//        seedinter = rr*seed*(65535-seed) ;
//        avg = seedinter[15:0];
        avg_r = seed;
    end
//    if xt == sometransformation of seed begin
//    avg <= xt;
//end
    else begin
//        #1;
        avg_r <= inter[15:0];
    end
end
endmodule
```

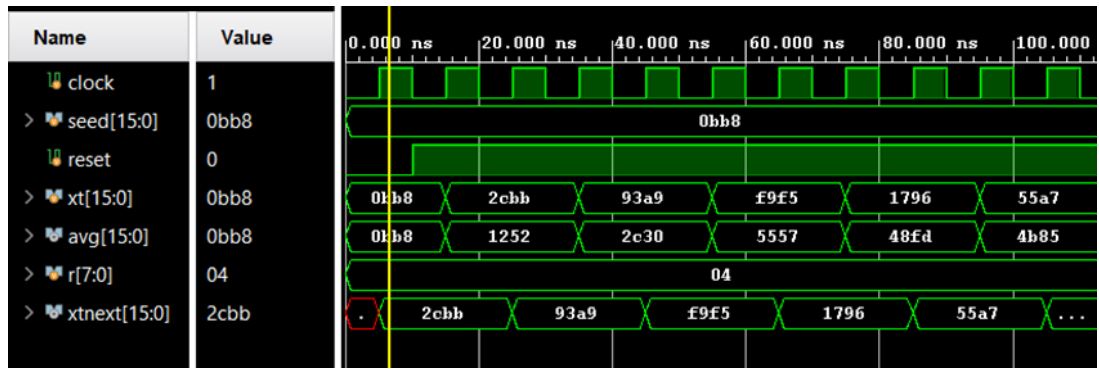

*Figure 2: Simulation for PRNG (Output PRN is called "avg" in the simulation).*

Essentially, we created the functions, ensuring we were using the correct nets, reg, and wire for each variable and that all the outputs and buffers were assigned correctly. We simulated this to ensure that our logic for the modules was correct before integrating them with the other modules meant for XADC, UART, clock, and 7-segment display.

## FPGA Setup

We decided on an overall system architecture, taking a top-down approach to designing this PRNG to ensure an organised implementation. Then, we developed each module according to the system architecture, as depicted in Figure 3.

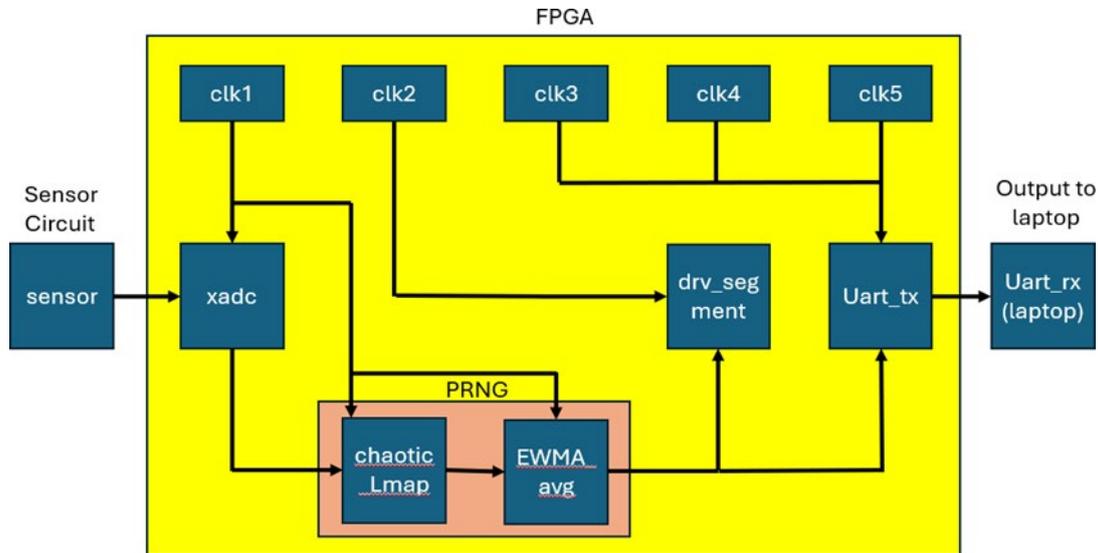
*Figure 3: Simplified System Architecture for our overall FPGA implementation.*

Each module has a specified purpose and is necessary for the functions of our PRNG.
- **xadc**: A module that reads and effectively outputs the readings obtained from the sensor, which are displayed in 16 bits.
- **chaotic_Lmap**: This module takes in a seed or initial value from the xadc and iteratively passes the value through the formula for a logistic map. This effectively creates a random variable, as the output is chaotic.
- **EWMA_avg**: A module used to take in a random value from the chaotic_Lmap and process it through the EWMA algorithm, outputting a 16-bit number that is effectively random and follows a Gaussian distribution.
- **drv_segment**: A module that takes in the random number to be displayed and displays it as a hex number onto a 7-segment display with 4 digits.
- **Uart_tx**: A module that transmits the 16-bit number in chunks of 8-bits to a receiver.
- **clk1**: A 1 Hz clock used for the updating of different values for the seed derived from the xadc module, as well as to update the different variables used in chaotic_Lmap and EWMA_avg.
- **clk2**: The module drv_segment drives the 500 Hz clock cycle between the different digits of the 7-segment display.
- **clk3**, **clk4**, **clk5**: Three clocks of different frequencies are essential to the functions of the UART. They cycle between chunks of 8 bits to transmit and set the baud rate for serial communication to the laptop receiver.

For the sensor circuit, we used a microphone so that we could take in a seed that is dependent on the volume of the surroundings. The sensor can be swapped easily with other sensors, as the whole PRNG depends only on the internal algorithm.

# V. Actual Verilog HDL implementation

In this section, we will document the more interesting parts of the code used in the Verilog HDL implementation and the code used to receive and process the data on the receiver side (laptop).

For the PRNG algorithm, we kept it the same as in Figure 2, but the important things to note here are that for the chaotic_Lmap algorithm, we need its output to feedback to its input, but in Verilog, this gets a bit tricky. A quick workaround for this is having an always loop running in the top module, to always update the input to the chaotic_Lmap based on its output. Also, note that we use a wire called "pee" to feed the output of the chaotic_Lmap to the input of the EWMA_avg.

```verilog
module top_module ();
…

//only register when button is pressed rstn
always@(negedge rstn or posedge CLK1Hz)begin
    if(!rstn)begin
        xnext <= seed;
        displayed_number_r <= 16'b0;
        end
    else begin
        xnext <= pee;
        displayed_number_r <= disp;
    end
end

chaotic_Lmap Lmap_u0(CLK1Hz, xnext, r, pee);

…
endmodule
```

For the XADC portion in the top_module, inside the always@(posedge CLK1Hz) begin loop, we take the data acquired from the microphone, assign it to the seed_r variable and let it read data every 1 second. This seed_r variable will then be used to initialise the PRNG.
We also changed the .daddr_in(PIN16_ADDR) to accommodate the microphone data pin.

```verilog
always @(posedge CLK1Hz) begin
//    disp <= ADC_data;
    if (ADC_data == 16'd0) begin
        seed_r <= 16'd1;
        end
    else begin
        seed_r <= ADC_data;
    end
end
```

```verilog
xadc_wiz_0 xadc_u0
(
    .daddr_in(PIN16_ADDR),      // Address bus for the dynamic reconfiguration port
    .dclk_in(sysclk),           // Clock input for the dynamic reconfiguration port
    .den_in(enable),            // Enable Signal for the dynamic reconfiguration port
```

For the top_module, note that we set the seed's default value to 1 instead of 0. If the seed input to the chaotic_Lmap is 0, the sequence could potentially collapse, and all the outputs from then on will keep on being 0.

For the uart_tx portion in the top_module, we removed the disp_r <= odd ? disp_r + 1'b1 : disp_r; as it generated too many numbers that did not fit the Gaussian Distribution we were aiming for. We also flipped the order in the disp_r <= odd ? disp_r + 1'b1 : disp_r; the numbers would come out as reflected on the 7 segment.

```verilog
always @(negedge rstn2 or posedge clk_data_proc) begin
    if (!rstn2) begin
        disp_r <= 8'h00;
        uart_tx_data <= 8'h00;
        odd <= 1'b0;
    end else begin
        disp_r <= disp;
        odd <= ~odd;
        uart_tx_data <= odd ? disp_r[15:8] : disp_r[7:0];
    end
end
```

```
12.
```

For the UART receiver function, which will be run on a laptop, we run a simple algorithm that stores each new updated value displayed on the 7-segment display. We use ser.read(2) to process 16 bits of data before running the int.from_bytes function, which changes the 16-bit number to an integer. Apart from that, we plot a histogram based on the stored values, and we overlay a normal distribution that uses the mean and standard deviation of the data. This allows us to see whether the output random numbers are Gaussian easily. The core Python code is shown in the text below.

```python
1.  def main():
2.      try:
3.          # Initialize serial connection
4.          ser = serial.Serial(port=serial_port,
5.                              baudrate=9600,
6.                              parity=parity,
7.                              stopbits=stop_bits,
8.                              bytesize=bytesize,
9.                              timeout=1)
10.         # Receive 100000 Byte
11.         prev_rx = 0
12.         data_cnter = [0] * 256
13.         for i in range(15000):
14.             rx_data = int.from_bytes(ser.read(2), byteorder='little')
15.             if rx_data != prev_rx:
16.                 data_cnter.append(rx_data)
17.                 prev_rx = rx_data
18.
19.             #data_cnter[rx_data] += 1
20.             #print(rx_data)
21.         print(data_cnter)
22.
23.         # Plot the histogram of the data
24.         n, bins, patches = plt.hist(data_cnter, bins=10, density=True, alpha=0.6, color='g')
25.
26. # Calculate the mean and standard deviation of the data
27.         mu, std = np.mean(data_cnter), np.std(data_cnter)
28. # Create a range of values (x) from the minimum to maximum bin edges
29.         xmin, xmax = plt.xlim()
30.         x = np.linspace(xmin, xmax, num=1000)
31.
32. # Calculate the normal distribution values for each x
33.         p = norm.pdf(x, mu, std)
34.
35. # Plot the normal distribution curve
36.         plt.plot(x, p, 'k', linewidth=4)
37.         plt.show()
```

## VI. Results

In the results section of our project, we demonstrate the effective operation of our pseudo-random number generator (PRNG) implemented on the FPGA. The histograms of the PRNG outputs, which were analysed both on the FPGA and via a connected laptop, consistently displayed a Gaussian distribution. This alignment with a Gaussian model confirms the theoretical predictions and the successful application of the Central Limit Theorem (CLT) to the outputs derived from the logistic map.

The figure below illustrates the distribution over roughly 60 iterations, which went on for 5 minutes of run time in real life. It shows the close fit between the empirical data and the Gaussian curve, thereby validating our approach. This evidence underscores the PRNG's capability to generate highly unpredictable and statistically sound sequences.

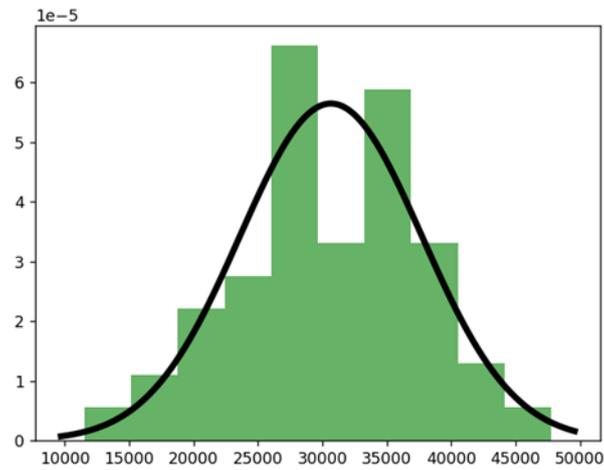
*Figure 4: Distribution of data received from CMOD XADC to PC via UART.*

Although we note achieving a nice fit of the pseudo-random numbers with a Gaussian curve, this fit would be even better if we had the time and resources to run the algorithm for hours, such that it would be a few thousand iterations. This would likely result in a smoother histogram that resembles a Gaussian curve.

We also put in some additional effort that included the design and 3D printing of a case for our whole setup. The mic sensor will be put inside the cone, and a phone will be set up, so its speaker is also within the cone. The cone will amplify/concentrate the sound and the mic sensor will be picking up on this amplified sound. Hence, the casing plays 2 roles: 1 to amplify the sound from a phone speaker and another to be used as the seed for our PRNG.

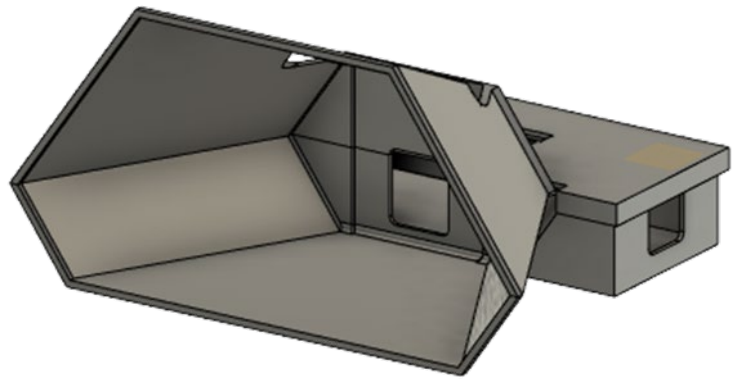
*Figure 5: 3D Model of Conical Structure to contain the microphone and CMOD circuitry.*

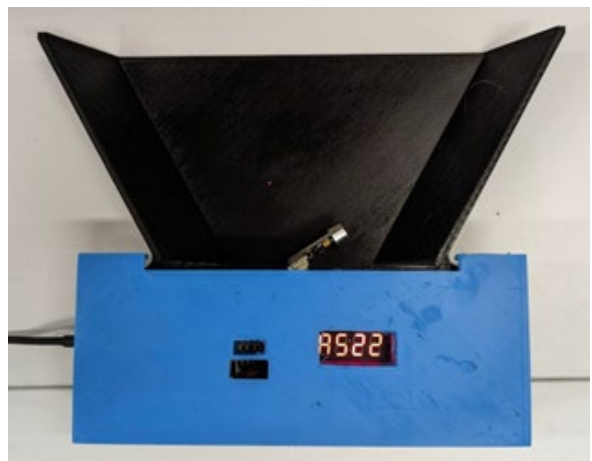
*Figure 6: Actual Conical structure containing the CMOD A7 and microphone.*

# VII. Acknowledgments

We would like to thank SUTD-ZJU IDEA Visiting Professor Grant (SUTD-ZJU (VP) 202103, and SUTD-ZJU Thematic Research Grant (SUTD-ZJU (TR) 202204), for supporting this work.